\documentclass[12pt,titlepage]{article}

\usepackage{float}
\newcommand{\be}{\begin{equation}}
\newcommand{\ee}{\end{equation}}

\newcommand{\bi}{\begin{itemize}}
\newcommand{\ei}{\end{itemize}}

\newif\ifdraft
\drafttrue 

\usepackage[pdftex]{graphicx}
\usepackage{amsmath}
\usepackage{textcomp}
\usepackage{caption}
\DeclareGraphicsExtensions{.jpg,.pdf,.mps,.png}

\usepackage{helvet}
\usepackage[]{amsmath}

\usepackage{cite}
\let\citeleft=(
\let\citeright=)

\graphicspath{{images/}}

\begin{document}
\bibliographystyle{mrm}

\begin{titlepage}

\begin{center}
	\begin{Large}
		\begin{bf}
Unraveling the Effect of Spatial Resolution and Scan Acceleration on 3D  Image-Based Navigators for Respiratory Motion Tracking in Coronary MR Angiography
\\ [0.1in]
		\end{bf}
	\end{Large}
\end{center}
\bigskip
\begin{center}
AUTHORS - Srivathsan P. Koundinyan$^1$, Joseph Y. Cheng$^1$, Mario O. Malav\'e$^1$, Phillip C. Yang$^3$, Bob S. Hu$^{4}$, Dwight G. Nishimura$^1$, Corey A. Baron$^2$

\end{center}
\vspace*{0.1in}
\noindent 1. Department of Electrical Engineering,
Stanford University, Stanford, California.  

\noindent 2. Department of Medical Biophysics, 
Western University, London, Ontario.  

\noindent 3. Department of Cardiovascular Medicine, 
Stanford University, Stanford, California.  

\noindent 4. Department of Cardiology, 
Palo Alto Medical Foundation, Palo Alto, California.

\noindent
{\em Running head:} \\
Unraveling the Effect of Spatial Resolution and Scan Acceleration on 3D Image-Based Navigators for Respiratory Motion Tracking in Coronary MR Angiography
 
\noindent

\noindent
{\em Address correspondence to:} \\
	Srivathsan P. Koundinyan \\
	Packard Electrical Engineering, Room 355 \\
	350 Serra Mall, Stanford, CA 94305-9510 \\
	Email: skoundin@stanford.edu
    
\noindent
This work was supported by NIH R01 HL127039, NIH T32HL007846, and GE Healthcare.

\noindent
Word Count: 250  (abstract)  3949  (body) 

\end{titlepage}

\section*{Abstract}
\setlength{\parindent}{0in}

Purpose: To study the accuracy of motion information extracted from beat-to-beat 3D image-based navigators (3D iNAVs) collected using a variable-density cones trajectory with different combinations of spatial resolutions and scan acceleration factors.

Methods: Fully sampled, breath-held 4.4 mm 3D iNAV datasets for six respiratory phases are acquired in a volunteer. Ground truth translational and nonrigid motion information is derived from these datasets. Subsequently, the motion estimates from synthesized undersampled 3D iNAVs with isotropic spatial resolutions of 4.4 mm (acceleration factor = 10.9), 5.4 mm (acceleration factor = 7.2), 6.4 mm (acceleration factor = 4.2), and 7.8 mm (acceleration factor = 2.9) are assessed against the ground truth information. The undersampled 3D iNAV configuration with the highest accuracy motion estimates in simulation is then compared with the originally proposed 4.4 mm undersampled 3D iNAV in six volunteer studies.  

Results: The simulations indicate that for navigators beyond certain scan acceleration factors, the accuracy of motion estimates is compromised due to errors from residual aliasing and blurring/smoothening effects following compressed sensing reconstruction. The 6.4 mm 3D iNAV achieves an acceptable spatial resolution with a small acceleration factor, resulting in the highest accuracy motion information among all assessed undersampled 3D iNAVs. Reader scores for six volunteer studies demonstrate superior coronary vessel sharpness when applying an autofocusing nonrigid correction technique using the 6.4 mm 3D iNAVs in place of 4.4 mm 3D iNAVs. 

Conclusion: Undersampled 6.4 mm 3D iNAVs enable motion tracking with improved accuracy relative to previously proposed undersampled 4.4 mm 3D iNAVs.


\setlength{\parindent}{0in}
{\bf Key words: 3D navigators, coronary angiography, motion correction}
\newpage

\section*{Introduction}
Motion is a major impediment to the acquisition of high quality images in free-breathing coronary magnetic resonance angiography (CMRA) exams \cite{ingle2014nonrigid, prieto2015highly, cruz2017highly}. A variety of retrospective motion correction methods have been proposed for CMRA. For these approaches to be effective, accurate motion measurements must be obtained. Many techniques have been developed to acquire this information, including 1D navigators placed over the diaphragm and self-navigation schemes that derive the position of the heart from the imaging data \cite{ehman1989adaptive, feng2016xd}. Another class of methods collects motion information using separately acquired 2D images of the heart \cite{wu2013free, correia2018optimized, malave2019whole, bustin2019five}.  

3D image-based navigators (3D iNAVs) have been proposed in recent years to directly monitor nonrigid motion in different regions of the heart \cite{keegan2007non, moghari2014three, powell2014cmra, addy20173d}. The premise underlying these approaches is to rapidly acquire a low-resolution 3D dataset in each cardiac cycle concurrently with the segmented high-resolution data that contributes to the final image over several heartbeats. Moghari, \textit{et al}. initially collected such 3D iNAVs with an anisotropic resolution of 56 x 18 x 1 mm\textsuperscript{3} using a Cartesian trajectory \cite{moghari2014three}. Powell, \textit{et al}. extended this approach with parallel imaging to acquire 3D iNAVs exhibiting an anisotropic resolution of 5 x 5 x 10 mm\textsuperscript{3} \cite{powell2014cmra}. By applying compressed sensing based parallel imaging alongside a variable-density (VD) cones trajectory, Addy, \textit{et al}. demonstrated 3D iNAVs with 4.4 mm isotropic resolution  \cite{addy20173d}. To track the highly local deformations of coronary vessels, these prior works have gradually augmented the spatial resolution of 3D iNAVs by increasing the associated scan acceleration factor. The resulting aliasing has been mitigated with iterative reconstruction. Residual undersampling artifacts, however, can remain in the 3D iNAVs in the case of large scan acceleration factors. Such artifacts may detract from the benefits of monitoring motion using 3D iNAVs with enhanced spatial resolution. 

In this work, we investigate the fidelity of motion estimates derived from 3D iNAVs collected with the accelerated cones trajectory described above. Determining the \textit{in vivo} performance of navigators is difficult because there is no reliable ground truth that can concurrently be acquired in real-time. To address this issue, we first develop a simulation framework to capture the translational displacements and nonrigid deformations of the heart introduced by the respiration cycle. Using this framework, we then examine the influence of different 3D iNAV spatial resolutions, corresponding to varying levels of necessary scan acceleration in VD cones imaging, on the accuracy of the extracted motion information. Finally, mindful of simulation results, we present a modified 3D iNAV design strategy. \textit{In vivo} nonrigid motion-corrected CMRA outcomes are utilized to compare the motion tracking capability of the proposed 3D iNAV design with that of the previously proposed approach. 

\section*{Methods}

\subsection*{Imaging Data and 3D iNAV Acquisition}

Beat-to-beat 3D iNAVs for respiratory motion tracking are collected as part of the cardiac-triggered, free-breathing 3D CMRA sequence shown in Supporting Information Figure S1 \cite{wu2013free}. Within each heartbeat, a fat saturation module is applied before the desired trigger delay point. Immediately following this module, imaging data is collected using a 3D cones $k$-space trajectory (28x28x14 cm\textsuperscript{3} FOV, 1.2 mm spatial resolution) \cite{gurney2006design}. An alternating-TR balanced steady state free precession (ATR bSSFP) readout is incorporated for further fat suppression and high blood signal. The overall acquisition scheme involves 9137 total interleaved cones acquired in segments of 18 every cardiac cycle. 

\begin{figure}
  \centering
    \includegraphics[width=\linewidth]{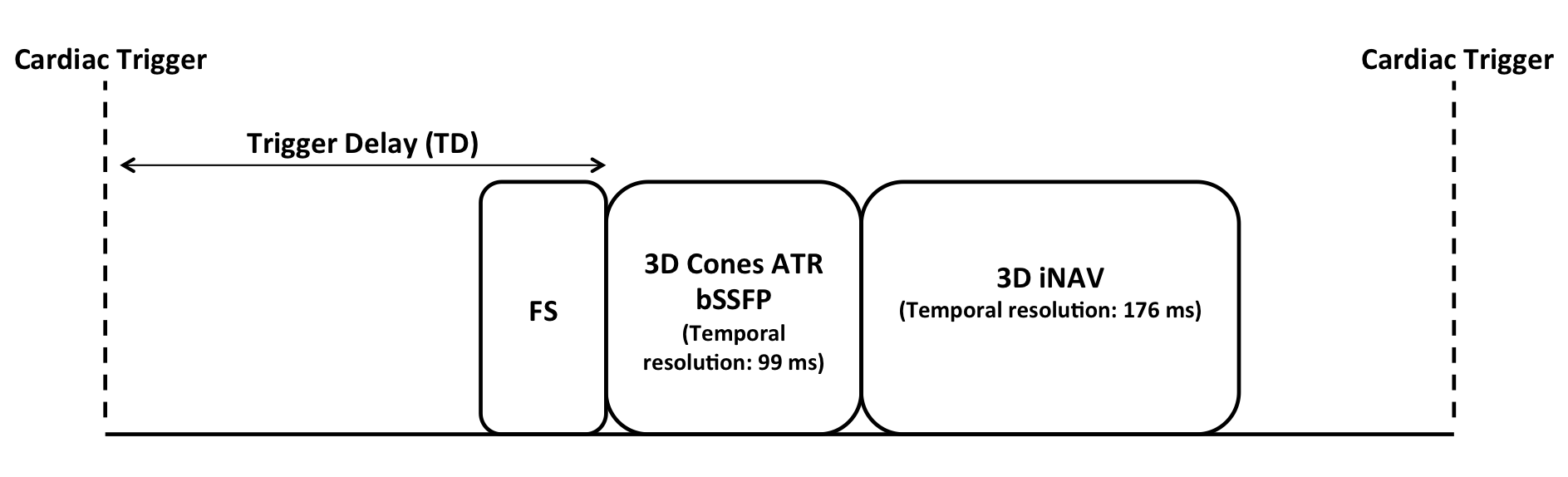}
  \caption*
    {Supporting Information Figure S1: Imaging data is collected with a cardiac-triggered sequence. A fat-saturation (FS) module is followed by a 3D cones sequence, where cones interleaves are acquired in groups of 18 with a temporal resolution of 99 ms during diastole. A 3D iNAV is also collected with a temporal resolution of 176 ms each heartbeat to monitor heart motion.
    }
\end{figure}

To collect a 3D iNAV in a single heartbeat, scan acceleration is applied with a VD cones trajectory \cite{addy2015high}. Here, the sampling density ($f$) in $k$-space ($|\boldsymbol{k}|$) is modified in the following manner:
\begin{equation} 
f(|\boldsymbol{k}|) = 
\begin{cases} 
      f_1 & |\boldsymbol{k}| \in [0, k_1] \\
      (f_1 - f_2)(1 - (|\boldsymbol{k}| - k_1)/(k_{max} - k_1))^{p} + f_2 &  |\boldsymbol{k}| \in (k_1, k_{max}] \\
   \end{cases}
\end{equation}
where the constant $f_1$ denotes the sampling density from 0 cm\textsuperscript{-1} to $k_1$ cm\textsuperscript{-1}. The transition in sampling density, from $f_1$ at $k_1$ down to $f_2$ at the maximum $k$-space extent ($k_{max}$), is governed by a $p$\textsuperscript{th} order polynomial. To mitigate undersampling artifacts in each 3D iNAV, reconstruction is performed with compressing sensing based parallel imaging using the state-of-the-art L\textsubscript{1}-ESPIRiT \cite{uecker2014espirit}:
\begin{equation} \label{eq:problemOld}
\underset{m}{\arg\min}\left\|DSm - y\right\|_2^2+\mu\left\|\Psi(m)\right\|_1
\end{equation}
where $D$ is the NUFFT operator, $S$ contains the coil sensitivity maps, $m$ is the desired 3D iNAV, $y$ is the acquired non-Cartesian data, $\mu$ is the regularization parameter, and $\Psi$ is the wavelet transform.

\begin{figure}
  \centering
    \includegraphics[width=\linewidth]{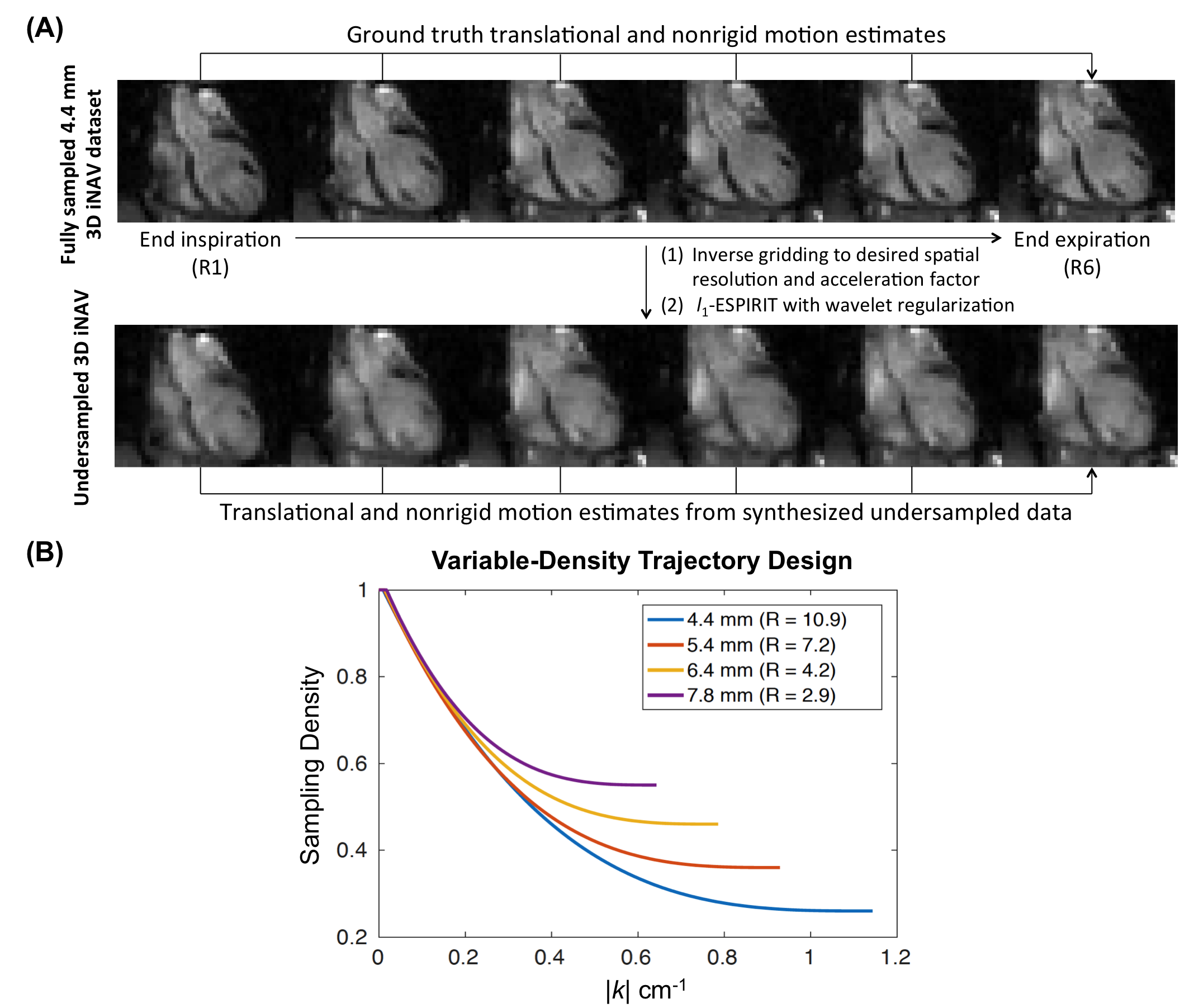}
  \caption[]
    {(A) Our simulation framework begins with six different respiratory phases acquired in separate breath-holds using a fully sampled 4.4 mm cones trajectory. Motion information from this dataset is used as the ground truth. Different undersampled 3D iNAV configurations are synthesized from the fully sampled navigators, and the corresponding motion information is compared with the ground truth to determine the optimal design for the 3D iNAVs. (B) We investigate 3D iNAVs with four spatial resolutions. Acceleration factors (R) for different spatial resolutions with a variable-density cones sampling pattern range from 10.9 to 2.9. 
    }
\end{figure}

\subsection*{Simulations}

We previously developed 4.4 mm isotropic spatial resolution 3D iNAVs using a 32-readout VD cones trajectory generated with $f_1$  = 1 (corresponding to a fully sampled region), $k_1$ = 1.14 m\textsuperscript{-1} (1\% of $k_{max}$),  $f_2$ = 0.27, and $p$ = 3.1. This corresponds to an acceleration factor of 10.9 compared to a fully sampled 3D cones trajectory requiring 349 readouts for 4.4 mm isotropic resolution. Prior work has demonstrated, however, that such a large acceleration factor when using an eight-channel coil can lead to blurring/smoothening effects or residual aliasing in compressed sensing reconstructions \cite{jaspan2015compressed}. The simulation framework described below aims to examine the influence of such reconstruction artifacts on the fidelity of motion estimates from 3D iNAVs. It additionally investigates potential benefits from reducing the spatial resolution of 3D iNAVs. This would decrease the necessary acceleration factor and thereby enhance the quality of compressed sensing reconstructions, which could improve the accuracy of the derived motion information. 

To develop the simulation framework, a volunteer was instructed to perform breath-holds in six respiratory phases from end-expiration to end-inspiration. For each respiratory phase, a fully sampled 4.4 mm spatial resolution 3D iNAV (349 readouts, TR = 5.5 ms, TE = 0.6 ms) was acquired with cardiac-gating (18 readouts per heartbeat, 99 ms temporal resolution) across 20 heartbeats. Because these six 3D iNAVs do not exhibit undersampling artifacts, the translational and nonrigid motion estimates obtained from them with respect to the end-expiration navigator serve as the ground truth (Figure 1(A)). Specifically, for each 3D iNAV, we first obtain 3D translational motion estimates. Rigid-body registration is performed with the MATLAB Image Processing Toolbox (The Mathworks, Natick, MA). Here, a 3D ellipsoidal mask covering the whole heart is prescribed and the mean-squared error within this mask is the similarity metric for registration. After the estimation of 3D translations and rigid-body alignment of 3D iNAVs, residual nonrigid motion in different respiratory phases is quantified using deformation fields, which are determined from diffeomorphic demons \cite{vercauteren2009diffeomorphic}. The choice of registration techniques for the estimation of translational and nonrigid motion mirrors those applied in the most recent work for processing 3D iNAVs \cite{luo2017nonrigid}. 

To analyze the effect of spatial resolution and undersampling, we generate a total of four VD cones trajectories with spatial resolutions of 4.4 mm, 5.4 mm, 6.4 mm, and 7.8 mm. In designing these trajectories, we fix the following parameters: FOV = 28x28x14 cm\textsuperscript{3}, number of readouts = 32, $f_1$  = 1, $p$ = 3.1, and $k_1$ = 1.14 m\textsuperscript{-1}. $k_{max}$ is prescribed to provide the desired spatial resolution and $f_2$ is modified to ensure each trajectory has 32 readouts (Figure 1(B)). Note that as the resolution of the trajectory decreases, $f_2$ (the sampling density in $k$-space periphery) increases, which results in smaller acceleration factors. 

$k$-space data for the different undersampled trajectories are computed from the fully sampled 4.4 mm data using a type 1 NUFFT (i.e., inverse gridding) (Figure 1(A)). Separate sensitivity maps for each respiratory phase are determined from the corresponding fully sampled acquisitions. These sensitivity maps are then used in the inverse gridding operation to generate multichannel $k$-space data for each respiratory phase. Following the synthesis of undersampled 3D iNAV data at varying spatial resolutions, reconstruction is performed with L\textsubscript{1}-ESPIRiT. For all reconstructions, the optimal regularization parameter is determined via a coarse-to-fine grid-based search that results in the lowest root-mean-squared error relative to the fully sampled image from which the non-Cartesian data was synthesized.

For each of the four undersampled 3D iNAV configurations, translational and nonrigid motion estimates relative to the end-expiration phase are computed using the aforementioned approach. Note that prior to deriving motion information, zero-padding of $k$-space is performed so that all 3D iNAV configurations have an interpolated isotropic spatial resolution of 4.4 mm. 

To assess errors in translations, we evaluate the absolute difference in superior-inferior (SI), anterior-posterior (AP), and right-left (RL) displacements derived from any particular 3D iNAV configuration relative to displacements from the fully sampled 4.4 mm 3D iNAV. The voxel-by-voxel SI, AP, and RL components of the nonrigid deformation fields from the undersampled 3D iNAVs are analyzed in a similar fashion. As an example, to compare the SI components of the deformation fields corresponding to the end-inspiration phase between an undersampled 3D iNAV and the ground truth 3D iNAV, we first compute the voxel-by-voxel absolute difference. Then, we examine the mean absolute difference in SI estimates across voxels spanning the heart (as determined by an ellipsoidal mask). The same procedure is carried out for the AP and RL components, and the overall error analysis is repeated for deformation fields associated with the remaining respiratory phases. 

Beyond individually analyzing the nonrigid SI, AP, and RL components, we consider the voxel-by-voxel error magnitude as well. For a voxel, this is defined as the square root of the sum of squares of the voxel-level errors in SI, AP, and RL directions. For each undersampled 3D iNAV configuration, the mean error magnitude in voxels spanning the heart is computed for the different respiratory phases. 

Inaccuracies in nonrigid motion estimates from the synthesized undersampled data can be due to the (1) lower spatial resolution with respect to the ground truth or (2) blurring and residual aliasing following iterative reconstruction. To separate these two effects, we create fully sampled 5.4 mm, 6.4 mm, and 7.8 mm datasets by appropriately truncating the \textit{k}-space data of the fully sampled 4.4 mm 3D iNAVs. The mean error magnitude relative to the fully sampled 4.4 mm 3D iNAVs is computed as described above for each of these generated datasets. By subtracting the error magnitude due to spatial resolution calculated here from the total error magnitude determined above, we isolate the influence of reconstruction artifacts in the 3D iNAVs on the extracted nonrigid motion information. 

\section*{Experiments}

The undersampled 3D iNAV configuration that provided the highest accuracy motion estimates in simulation was compared with the previously applied 4.4 mm 3D iNAV design in six volunteer acquisitions. Each volunteer underwent two scans with the CMRA sequence shown in Supporting Information Figure S1. Respiratory motion in each scan was tracked with either the modified 3D iNAV configuration or the 4.4 mm 3D iNAV design. The order of the two scans was randomized. All scans were carried out on a 1.5 T whole-body GE scanner with maximum slew rate of 150 mT/m/ms and maximum gradient amplitude of 40 mT/m. Participants provided informed consent, and the institutional review board approved the complete scan protocol. The studies were performed using an eight-channel cardiac receive coil with cardiac triggering via a peripheral plethysmograph. A 3D cones trajectory with the following imaging parameters was utilized: TR = 5.5 ms; flip angle = 70\textdegree; bandwidth = 250 kHz; FOV = 28x28x14 cm\textsuperscript{3}, resolution = 1.2 mm isotropic. The total scan time across all subjects spanned 508 heartbeats and ranged from 7 to 10 minutes due to variations in heartrate.

An autofocusing motion correction framework utilizing both translational and nonrigid estimates was applied to evaluate the motion information from the two different 3D iNAVs \cite{luo2017nonrigid}. The first step in this scheme entails the estimation of 3D translational motion from the beat-to-beat 3D iNAVs and subsequent correction of imaging data with linear phase terms. Following this, residual nonrigid motion in the rigid-body aligned 3D iNAVs is quantified using the deformation fields from diffeomorphic demons. k-Means clustering is then performed to group pixels with similar deformation fields over time into 32 clusters. Averaging of the deformation fields in each cluster generates a total of 32 localized 3D translational motion trajectories. For each localized translational estimate, the appropriate linear phase modulation is applied, and a candidate motion compensated image is reconstructed. From this collection of motion compensated images, a localized gradient entropy metric is used to assemble the final image on a voxel-by-voxel basis.

Two board-certified cardiologists with experience in CMRA assessed variation in nonrigid motion correction outcomes using the 4.4 mm 3D iNAV and the modified 3D iNAV configuration. Thin-slab maximal intensity projection reformats of the right coronary artery (RCA) and left coronary artery (LCA) were generated with OsiriX (Pixmeo, Geneva, Switzerland). The two autofocusing reconstructions (one applying the 4.4 mm 3D iNAV and the other applying the modified 3D iNAV configuration) were randomized and presented together, and the blinded readers scored the proximal, medial, and distal segments of the RCA and LCA on a five-point scale: 5-Excellent, 4-Good, 3-Moderate, 2-Poor, 1-Non-diagnostic. Paired two-tailed Student's t-tests were applied to determine significance.

\begin{figure}
  \centering
    \includegraphics[width=\linewidth]{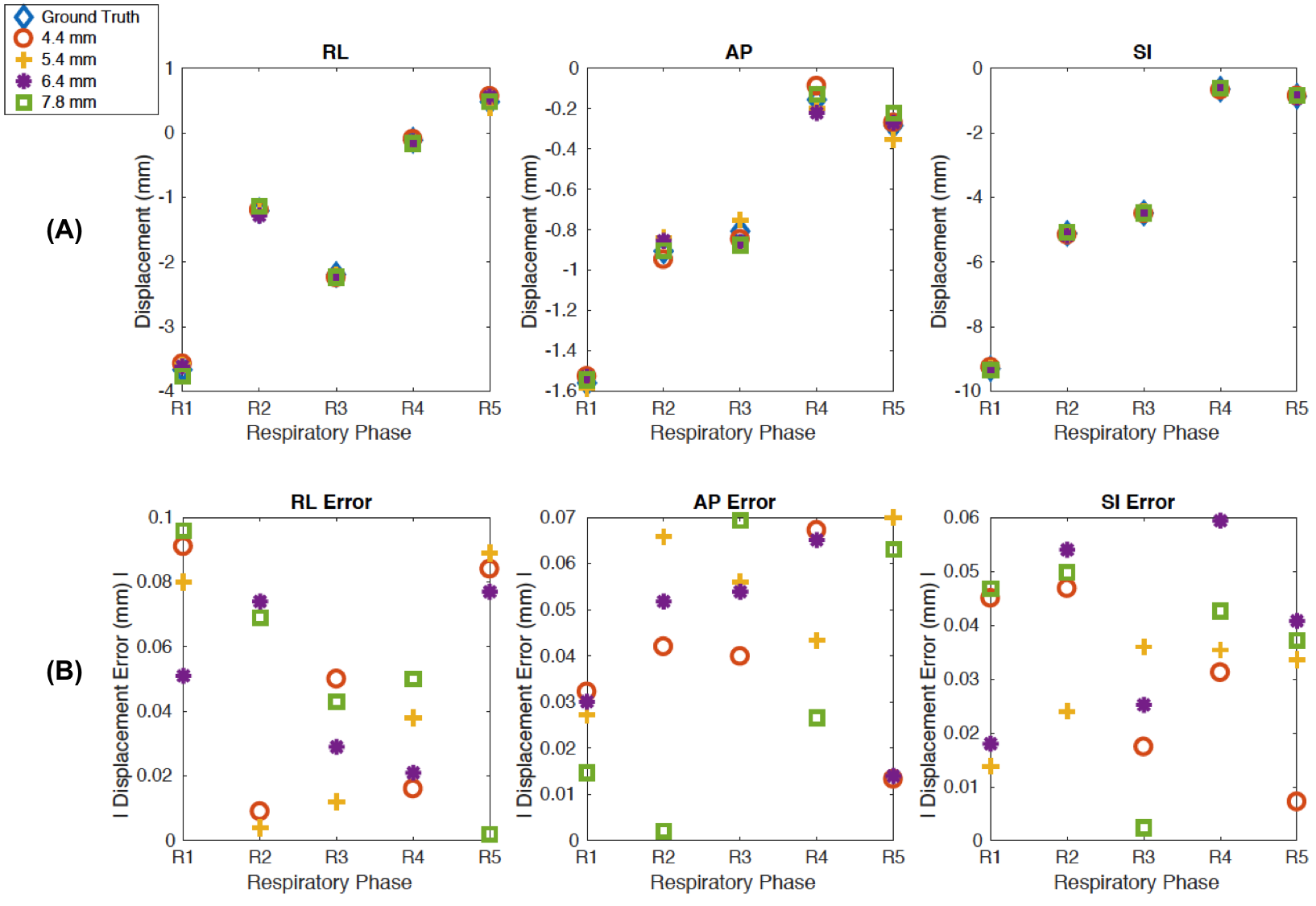}
  \caption[]
    {(A) The 3D translational displacements computed using the ground truth, fully sampled 4.4 mm 3D iNAV and the four undersampled 3D iNAVs exhibit similar trends across the respiratory phases (R1 = end-inspiration respiratory phase and R5 = respiratory phase closest to end-expiration). (B) The absolute difference in the RL, AP, and SI estimates from the undersampled 3D iNAVs relative to those from the fully sampled 3D iNAV present small errors below 0.1 mm. This indicates that all the undersampled 3D iNAVs are comparable in tracking the translational motion of the heart induced by respiration.
    }
\end{figure}

\begin{figure}
  \centering
    \includegraphics[width=\linewidth]{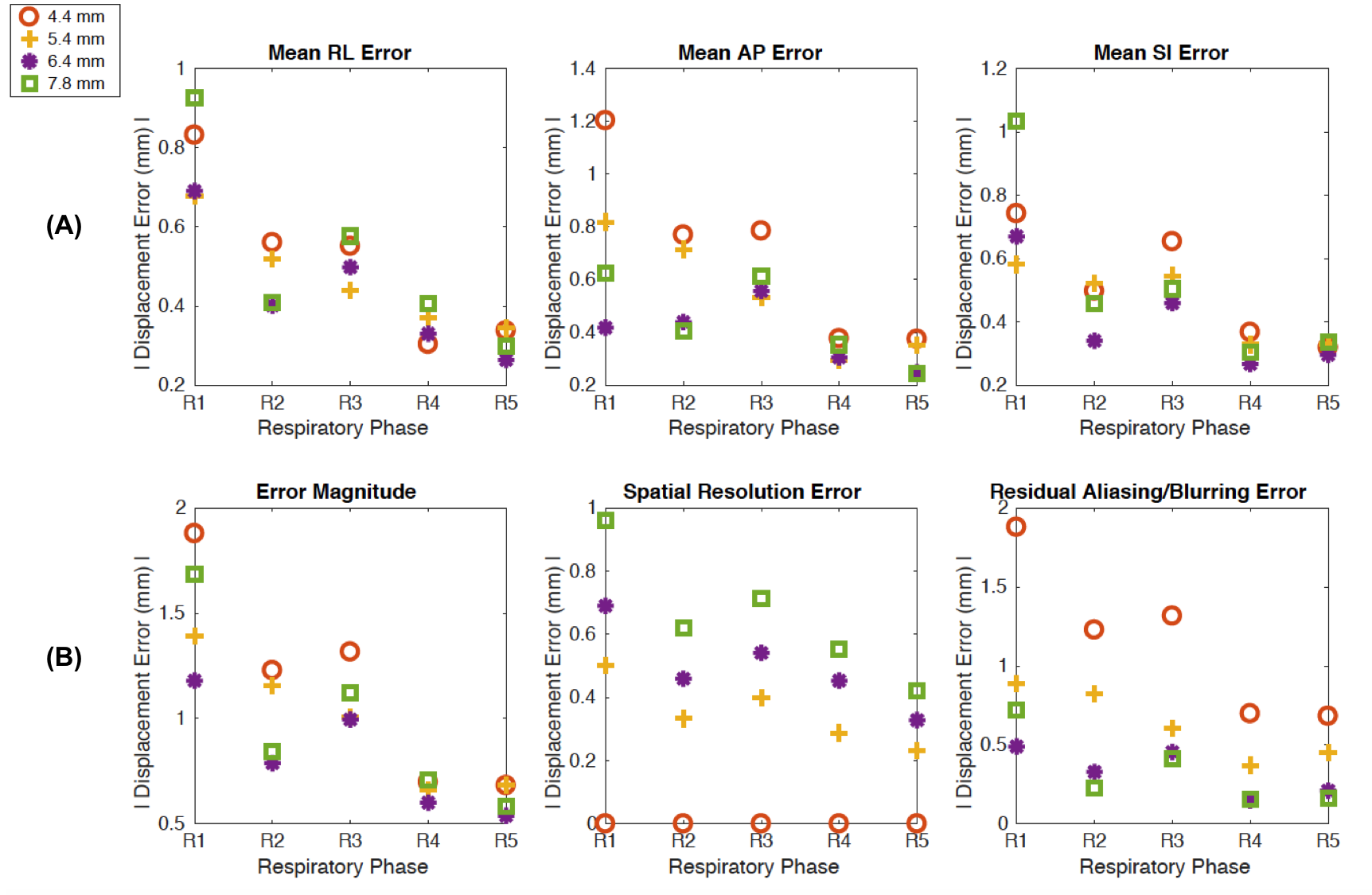}
  \caption[]
    {Compared to the inaccuracies observed in translational estimates, the mean errors (relative to the fully sampled 4.4 mm 3D iNAV) are larger in the RL, AP, and SI components of the nonrigid deformation fields from the four undersampled 3D iNAVs, as shown in (A) for different respiratory phases (R1 = end-inspiration respiratory phase and R5 = respiratory phase closest to end-expiration). With the exception of the RL error in respiratory phase R4, the 6.4 mm 3D iNAV consistently provides higher accuracy nonrigid motion estimates than the 4.4 mm 3D iNAV. This trend is more apparent in the error magnitude shown in (B) for the undersampled 3D iNAVs. Across all respiratory phases and undersampled 3D iNAV configurations, the 6.4 mm 3D iNAV results in the lowest errors. The errors due to spatial resolution (determined by comparing fully sampled navigators at different resolutions to the ground truth 4.4 mm 3D iNAV) and residual aliasing/blurring (computed by subtracting the spatial resolution error from the error magnitude) effects following L\textsubscript{1}-ESPIRiT combine to give the highest accuracy nonrigid estimates for the 6.4 mm 3D iNAV.
    }
\end{figure}

\section*{Results}

Figure 2(A) presents 3D translations computed with the fully sampled 3D iNAV and the four undersampled 3D iNAV configurations in simulation. Absolute differences in translations from the undersampled 3D iNAVs relative to those from the ground truth 3D iNAV are shown in Figure 2(B). As is evident, the undersampled 3D iNAVs perform similarly to one another for the estimation of translational displacements. Moreover, all the errors in translations are below 0.1 mm. 

The average errors in the nonrigid deformation fields from the different undersampled 3D iNAVs are highlighted in Figure 3(A). For the RL component, with the exception of one respiratory phase, the 6.4 mm 3D iNAV presents lower errors than the 4.4 mm 3D iNAV applied in prior work. A similar trend is observed for the AP component, where the 6.4 mm 3D iNAV consistently outperforms the 4.4 mm 3D iNAV. In the SI component, for four respiratory phases, the 6.4 mm 3D iNAV yields the smallest errors among the assessed 3D iNAV configurations. 

The mean error magnitude combining the individual RL, AP, and SI errors accentuates the observed patterns (Figure 3(B)). Here, across all the respiratory phases, the 6.4 mm 3D iNAV exhibits the lowest error magnitude. Note that in the case of fully sampled datasets at different resolutions, it is the 4.4 mm 3D iNAV that exhibits the lowest error magnitude. However, in the case of an undersampled 4.4 mm 3D iNAV with large scan acceleration (R = 10.9), the benefit from high spatial resolution for motion tracking is offset by the error contribution from residual aliasing and blurring/smoothening effects following L\textsubscript{1}-ESPIRiT. The simulation suggests that an undersampled 6.4 mm 3D iNAV (R = 4.2) balances the tradeoff between (1) improved motion tracking with high spatial resolution navigators and (2) compromised motion tracking in the presence of reconstruction artifacts due to aggressive scan acceleration. 

\begin{figure}
  \centering
    \includegraphics[width=\linewidth]{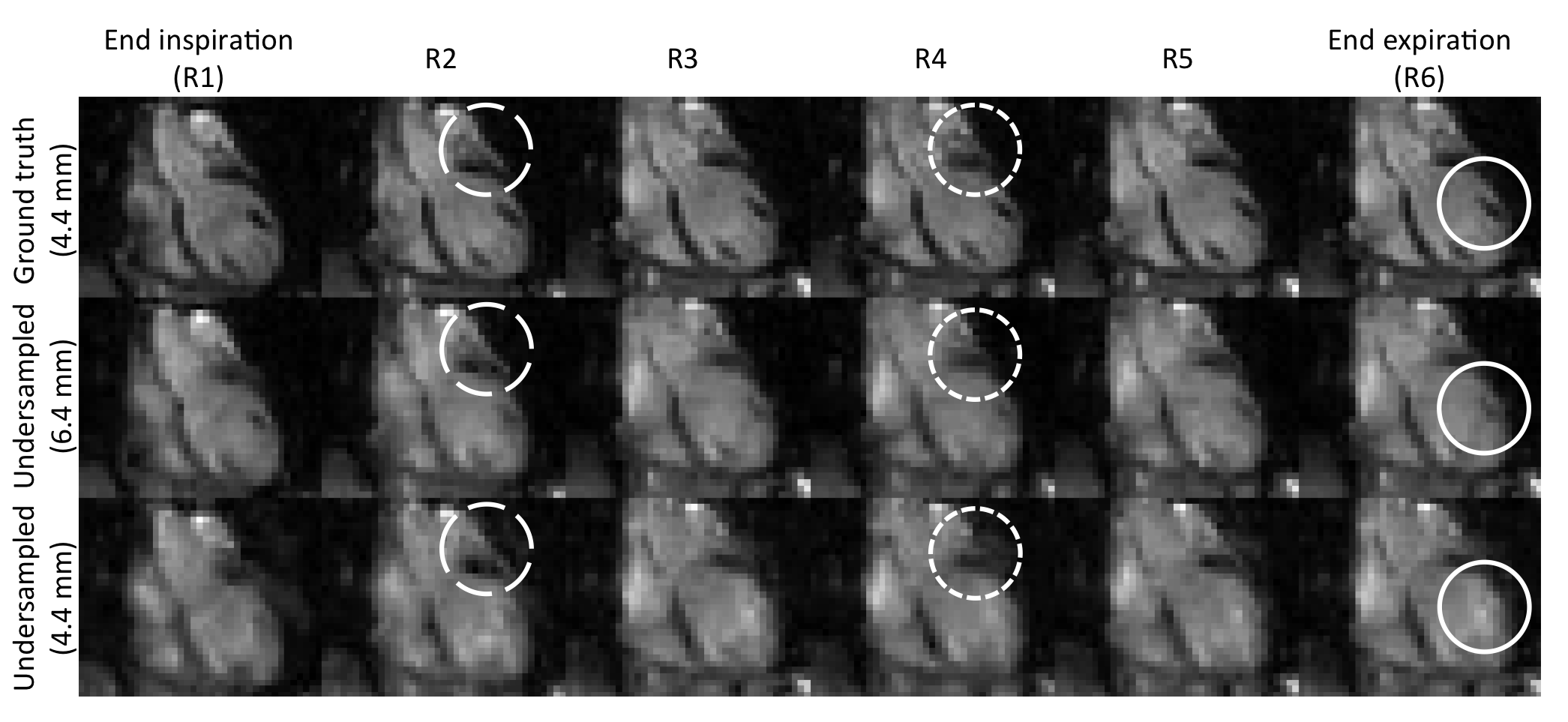}
  \caption*
    {Supporting Information Figure S2: The synthesized undersampled 4.4 mm 3D iNAV with an acceleration factor of 10.9 exhibits residual aliasing as well as blurring and smoothening effects that skew the quantification of nonrigid motion information. The undersampled 6.4 mm 3D iNAV requires a lower acceleration factor (4.2), which lessens the severity of artifacts following reconstruction. Dotted circles in the different respiratory phases highlight regions in which structure is similarly depicted in the ground truth 3D iNAV and undersampled 6.4 mm 3D iNAV, but poorly seen in the undersampled 4.4 mm 3D iNAV due to reconstruction artifacts. 
    }
\end{figure}

\begin{figure}
  \centering
    \includegraphics[width=\linewidth]{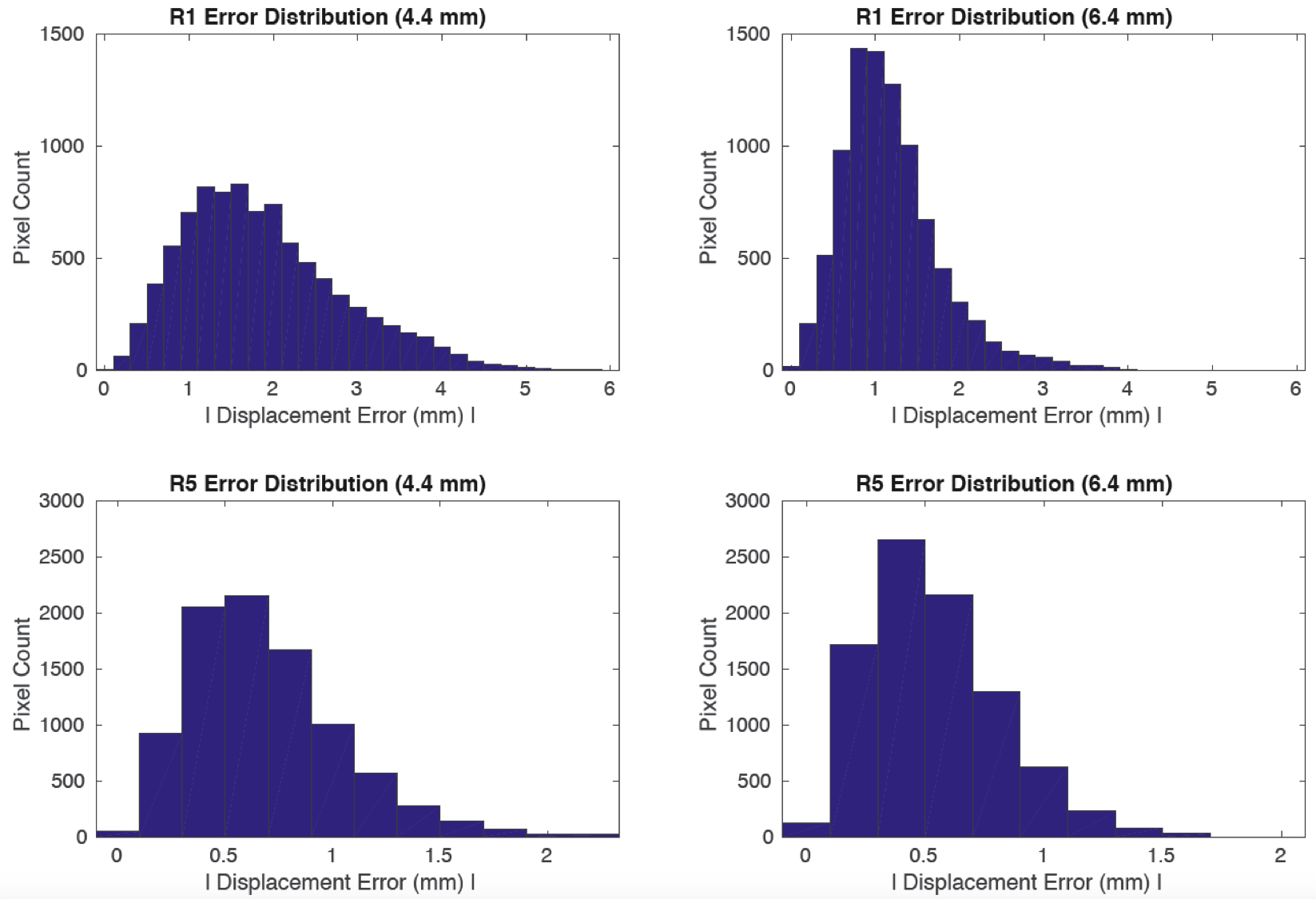}
  \caption*
    {Supporting Information Figure S3: The voxel-by-voxel distribution in error magnitude for two respiratory phases (R1 = end-inspiration respiratory phase and R5 = respiratory phase closest to end-expiration) demonstrate the advantages of using a lower resolution 3D iNAV. For both respiratory phases, the 6.4 mm 3D iNAV has larger pixel counts near smaller errors compared to the 4.4 mm 3D iNAV. This trend is accentuated in R1, which exhibits more nonrigid deformations than R5 since the end-expiration respiratory phase is the reference frame. 
    }
\end{figure}

\begin{figure}
  \centering
    \includegraphics[width=0.7\textwidth]{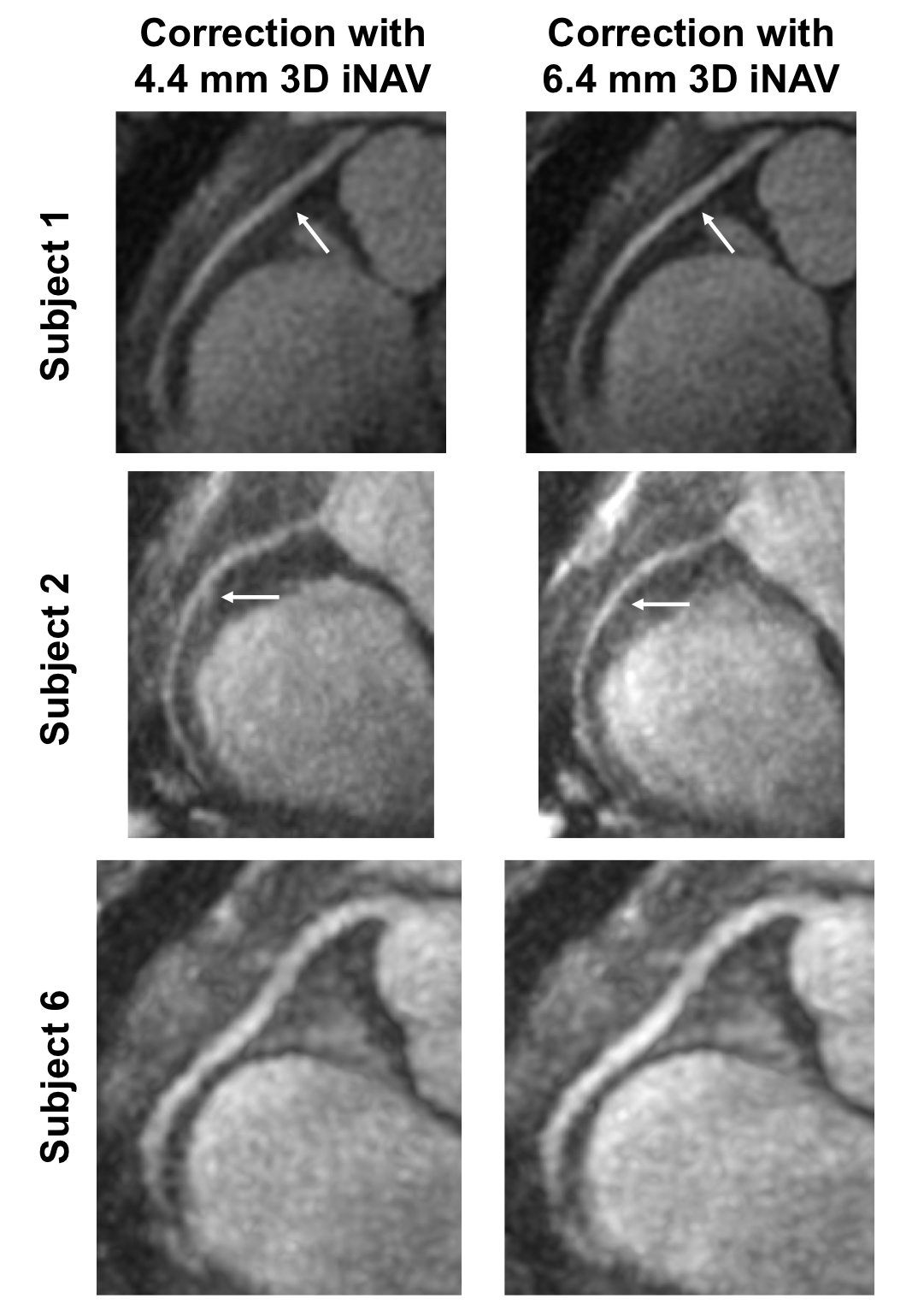}
  \caption[]
    {Reformatted MIP of the RCA with autofocusing motion correction using the 4.4 mm 3D iNAV (left) and 6.4 mm 3D iNAV (right) for three subject studies. In Subject 1 and Subject 2, applying the motion information from the 6.4 mm 3D iNAV better delineates the RCA compared to utilizing the 4.4 mm 3D iNAV. Subject 6 is a case where the depiction of the different RCA segments is similar between the two approaches. White arrows indicate regions of notable differences between autofocusing using the 4.4 mm 3D iNAV and 6.4 mm 3D iNAV.
    }
\end{figure}

\begin{figure}
  \centering
    \includegraphics[width=0.9\textwidth]{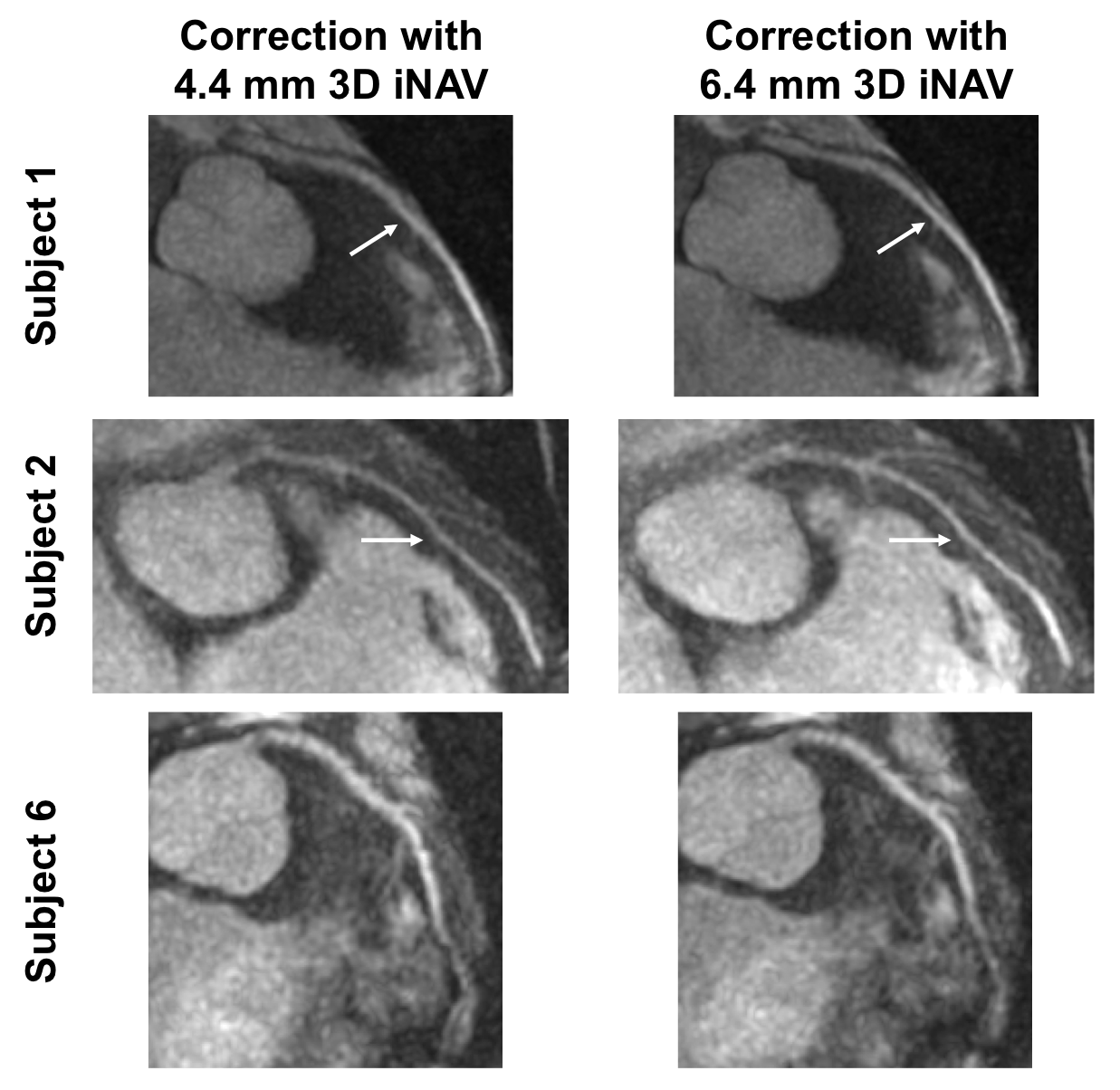}
  \caption[]
    {LCA with autofocusing motion correction using the 4.4 mm 3D iNAV (left) and 6.4 mm 3D iNAV (right). As seen with the RCA, the 6.4 mm 3D iNAV presents the LCA in an enhanced manner relative to the 4.4 mm 3D iNAV in Subject 1 and Subject 2. The LCA is visualized with equivalent detail in Subject 6 irrespective of the 3D iNAVs used for motion tracking. Differences in LCA sharpness are highlighted by white arrows. 
    }
\end{figure}

\begin{figure}
  \centering
    \includegraphics[width=\linewidth]{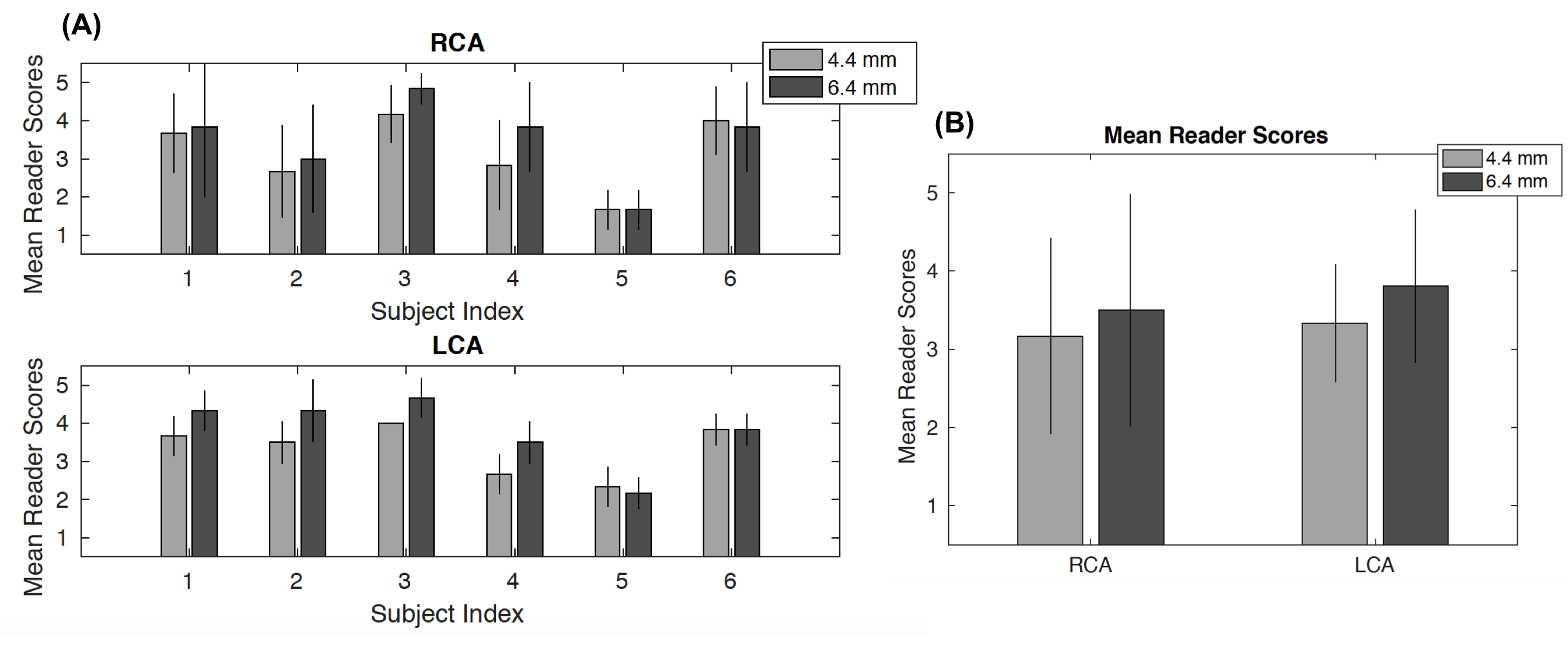}
  \caption[]
    {(A) The average scores of the RCA and LCA across both readers and three coronary segments in six volunteers when using the 4.4 mm 3D iNAV and 6.4 mm 3D iNAV for autofocusing. (B) The mean RCA and LCA reader scores across all volunteers for nonrigid correction outcomes when applying the two 3D iNAVs. The statistical significance of the results for reader scores was P $<$ 0.05 for the RCA as well as the LCA when using the two-tailed Student's t-test. 
    }
\end{figure}

Supporting Information Figure S2 presents all the respiratory phases for the fully sampled 4.4 mm 3D iNAV, undersampled 4.4 mm 3D iNAV, and undersampled 6.4 mm 3D iNAV. Relative to the undersampled 4.4 mm 3D iNAV, the undersampled 6.4 mm 3D iNAV exhibits improved depiction of various structures in the heart. To further substantiate the better performance of the 6.4 mm 3D iNAV compared to the 4.4 mm 3D iNAV in simulations, Supporting Information Figure S3 shows a histogram of the error magnitude in voxels spanning the heart for two sample respiratory phases. In both phases, the distribution of error magnitude is centered around smaller errors for the 6.4 mm 3D iNAV relative to the 4.4 mm 3D iNAV.  

The beat-to-beat 4.4 mm 3D iNAVs and 6.4 mm 3D iNAVs obtained from the six volunteer studies corroborate the trends seen in simulations (Supporting Information Video S1). Specifically, across all volunteers, the 6.4 mm 3D iNAVs exhibits less residual aliasing and blurring/smoothening artifacts following L\textsubscript{1}-ESPIRiT. For example, in Subject 1, the boundary between the apex of the heart and diaphragm as well as the separation of the ventricles of the heart is best delineated with the 6.4 mm 3D iNAVs. Such structural features can enhance the performance of nonrigid image registration techniques for deriving motion information \cite{sroubek2004registration, vandewalle2006frequency}. 

Reformatted oblique thin-slab MIP images depicting the RCA for three volunteer studies are shown in Figure 4. For Subject 1 and Subject 2, autofocusing correction with the 6.4 mm 3D iNAVs results in enhanced depiction of the medial RCA segment compared to correction with the 4.4 mm 3D iNAVs. In the case of Subject 6, the RCA exhibits equivalent sharpness regardless of the applied 3D iNAVs. Similar trends are observed for the LCA in Figure 5. While the medial segment of the LCA is better visualized when using the 6.4 mm 3D iNAVs in place of the 4.4 mm 3D iNAVs for Subject 1 and Subject 2, a difference in sharpness is not observed in the LCA of Subject 6. 

Figure 6 showcases the results of the qualitative reader studies. When using the 4.4 mm 3D iNAV for autofocusing correction, the average score for the RCA and LCA is 3.17 and 3.33, respectively, across both readers, all segments, and all subjects. Applying the 6.4 mm 3D iNAV results in scores of 3.50 and 3.81. Observed discrepancies in scores between motion correction with the 4.4 mm 3D iNAVs and 6.4 mm 3D iNAVs are statistically significant with P $<$ 0.05. The correlation coefficients between each reader for the RCA and LCA are 0.83 and 0.61, respectively. 


\section*{Discussion}

In this work, we examined the accuracy of translational and nonrigid motion estimates offered by undersampled 3D iNAVs acquired with a VD cones trajectory. This analysis was performed using a novel simulation framework for investigating motion estimation errors in CMRA. To obtain ground truth motion information, fully sampled, breath-held 4.4 mm 3D iNAV datasets were collected at several respiratory phases. Then, different undersampled 3D iNAV configurations with spatial resolutions ranging from 4.4 mm to 7.8 mm and scan acceleration factors between 10.9 and 2.2 were generated from the fully sampled data. While translational motion from the undersampled 3D iNAVs strongly correlated with those from the fully sampled 3D iNAVs, the nonrigid motion estimates exhibited large errors. Notably, we found that the undersampled 3D iNAV with the highest spatial resolution (4.4 mm) did not provide the best accuracy motion information. This is because the undersampled 4.4 mm 3D iNAV also has the largest associated scan acceleration factor (R = 10.9), leading to artifacts such as aliasing and blurring/smoothening effects after iterative reconstruction. We demonstrated that the 6.4 mm 3D iNAV, while lower in resolution, provides higher fidelity nonrigid motion estimates, as it achieves a sufficient spatial resolution with a moderate acceleration factor (R = 4.2). 

The simulation framework developed in this study does not incorporate several important considerations. First, the sensitivity maps used to generate multichannel, undersampled \textit{k}-space data are the same maps applied in L\textsubscript{1}-ESPIRiT to reconstruct 3D iNAVs. In practice, errors exist in the sensitivity maps for L\textsubscript{1}-ESPIRiT, which might worsen the accuracy of motion information from 3D iNAVs. Second, note that the 3D iNAVs are acquired as part of a bSSFP sequence. Thus, higher resolution 3D iNAVs will experience more eddy current artifacts, which we do not study with our simulations \cite{malave2019whole}. This suggests, however, that the improvements of the 6.4 mm 3D iNAVs over 4.4 mm 3D iNAVs are likely even larger than indicated by our simulations. Third, the effect of subject size on 3D iNAV quality is not evaluated in the current work. Specifically, while the VD cones trajectories for all undersampled 3D iNAVs are designed with a nominal FOV of 28x28x14 cm\textsuperscript{3}, the effective FOV is smaller for the 4.4 mm 3D iNAV than the 6.4 mm 3D iNAV. This is because of the 2.6x (= 10.2/4.2) greater scan acceleration factor for the 4.4 mm 3D iNAV compared to the 6.4 mm 3D iNAV. As a result, in the case of large subjects, the aliasing due to undersampling will be more pronounced for the 4.4 mm 3D iNAV relative to the 6.4 mm 3D iNAV. Consequently, the presence of reconstruction artifacts following L\textsubscript{1}-ESPIRiT might be more exaggerated in the higher resolution 3D iNAV. Further study is warranted to examine the influence of subject size on motion tracking accuracy with undersampled 3D iNAVs, as the simulation is based on fully sampled data from a single volunteer. Lastly, the ability of this volunteer to perform breath-hold at six respiratory phases could impact the conclusions from the simulation. However, as long as there are several unique respiratory positions, it should nevertheless serve as a sufficient proxy for evaluating the nonrigid motion estimation performance of different 3D iNAV configurations.

Despite the simplifications in our simulation, the findings from it correctly guided us in our experimentation. In the six volunteer studies, the 6.4 mm 3D iNAVs improved the depiction of cardiac structure compared to the 4.4 mm 3D iNAVs. Accordingly, nonrigid autofocusing correction of free-breathing CMRA data yielded sharper coronary vessels with the lower resolution 3D iNAVs. Assessment by two cardiologists validated these trends. 

The L\textsubscript{1}-ESPIRiT technique applied in this work utilizes spatial regularization alone. Temporal regularizers such as total variation or low rank constraints across the navigator frames might further mitigate aliasing artifacts. Temporal constraints may improve reconstruction quality, but their ability to retain motion information has not been studied in a quantitative manner. Therefore, we did not incorporate these additional regularizers in our approach to reconstruct 3D iNAVs. Note also that parallel imaging and compressing sensing reconstructions are performed here using an eight-channel cardiac coil. A larger number of channels will enable greater accelerations for acquiring 3D iNAVs. The simulation pipeline developed in this work can readily be applied in contexts involving coils with additional elements to understand the impact of scan acceleration on the derived motion information. By leveraging coils with several channels alongside reconstruction schemes with a combination of regularizers, 3D iNAVs can potentially be directly derived from the high-resolution imaging data. 

\section*{Conclusion}

We have analyzed the effect of spatial resolution and scan acceleration on the fidelity of respiratory motion tracking using 3D iNAVs for CMRA. Through simulations, we determined that a higher spatial resolution 3D iNAV, if fully sampled, results in better motion estimates. However, with undersampling, the advantages associated with high spatial resolution motion tracking are offset by the presence of artifacts following iterative reconstruction. In light of this, we found that an undersampled 4.4 mm 3D iNAV (R = 10.9) yielded lower accuracy nonrigid motion information than an undersampled 6.4 mm 3D iNAV (R = 4.2). \textit{In vivo} CMRA studies presenting sharp autofocusing motion correction outcomes demonstrated a capability for monitoring motion with improved fidelity using the 6.4 mm 3D iNAV in place of the 4.4 mm 3D iNAV.  

\newpage
\bibliography{refs_cbaron}

\end{document}